\documentclass[preprint]{aastex}

\shorttitle{}
\shortauthors{}

\begin{document}

\title{On the Formation of Massive Primordial Stars}
\author{Kazuyuki Omukai \altaffilmark{1,2}
 and Francesco Palla \altaffilmark{1}}
\altaffiltext{1}{Osservatorio Astrofisico di Arcetri, Largo E. Fermi 5,
50125 Firenze, Italy}
\altaffiltext{2}{Division of Theoretical Astrophysics, 
National Astronomical Observatory, Mitaka, Tokyo 181-8588, Japan}
\email{omukai@th.nao.ac.jp, palla@arcetri.astro.it}
\begin{abstract}
We investigate the formation by accretion of massive primordial protostars in
the range 10 to 300 $M_\odot$.  The high accretion rate used in the models
($\dot{M}=4.4 \times 10^{-3}\, M_{\sun}\, {\rm yr}^{-1}$) causes the
structure and evolution to differ significantly from those of both
present-day protostars and primordial zero-age main sequence stars. After an
initial expansion of the radius (for $M_\ast\la 12 M_\odot$), the protostar
undergoes an extended phase of contraction (up to $M_\ast \simeq 60 M_\odot$).
The stellar surface is not visible throughout most of the main accretion
phase, since a photosphere is formed in the infalling envelope. Also,
significant nuclear burning does not take place until a protostellar mass of
about $80\, M_{\sun}$. As the interior luminosity approaches the Eddington
luminosity, the protostellar radius rapidly expands, reaching a maximum
around $100 \,M_\odot$.  Changes in the ionization of the surface layers
induce a secondary phase of contraction, followed by a final swelling due to
radiation pressure when the stellar mass reaches about $300\, M_{\sun}$.
This expansion is likely to signal the end of the main accretion phase, thus
setting an upper limit to the protostellar mass formed in these conditions.
\end{abstract}

\keywords{cosmology: theory --- early universe  
--- stars: formation --- stars: pre-main-sequence}

\section{Introduction}
Numerical simulations are now beginning to constrain the properties 
of the sites where star formation first occurred in the post-recombination
epoch (e.g., Abel, Bryan, \& Norman 2000; Bromm, Coppi, \& Larson 1999, 2001;
Tsuribe 2001).  These works indicate that the fragment mass scale of
primordial gas clouds is relatively large ($M_f\sim 10^{3} M_{\sun}$).  On
the other hand, the mass of protostellar cores formed inside these fragments
is less than $10^{-2}M_{\sun}$ at the beginning of the main accretion phase
(Palla, Salpeter, \& Stahler 1983; Omukai \& Nishi 1998). Because of the
large reservoir, the cores continue to grow in mass owing to accretion of
ambient matter.  Eventually, they become ordinary stars that shine by
nuclear burning.  The fundamental questions are then: What fraction of the
cloud mass can be incorporated into a star? Is there an upper limit to the
mass of the primordial stars forming under such circumstances?

In the context of present-day star formation, it is generally believed that
stellar activity (radiation force, the expansion of an HII region, stellar
winds, etc.) of sufficiently massive protostars halts accretion, thereby
determining the final stellar mass (e.g., Larson \& Starrfield 1971; Nakano,
Hasegawa, \& Norman 1995).  Of these processes, radiation force on dust grains
appears to be the most efficient, making spherical accretion above a mass of
$\sim 15\, M_{\sun}$ increasingly difficult for solar-type metallicities
(e.g., Wolfire \& Cassinelli 1987), 
although non-spherical accretion may enhance the maximum stellar mass 
(Nakano 1989).

The main differences between primordial and Pop I accreting protostars are
the following.  First, the opacity of the infalling gas is
drastically reduced due to the lack of dust grains.  Second, the mass
accretion rate, $\dot{M}$, which is related to the sound speed $c_{\rm s}$ in
the protostellar cloud by $\dot{M}=c_{\rm s}^{3}/G$ (e.g., Stahler, Shu \&
Taam 1980),
is substantially higher, because of the high temperature of the primordial
clouds, typically $T_g\sim 1000-1500$~K  over a wide density range (Palla et
al. 1983).  These two properties imply a reduced effect of radiation pressure
of the protostellar photons and a higher momentum of the inflow.  Thus,
reversal of accretion by radiation force alone seems to be less efficient, and
a higher maximum stellar mass is expected (e.g., Larson \& Starrfield 1971).
Although qualitatively correct, this prediction has not been substantiated by
numerical calculations covering a large mass interval. We still do not know
the sequence of events that lead to the termination of the infall at the
onset of protostellar surface activity. It is the purpose of this Letter
to illustrate the results of such calculations for accreting protostars in
the mass range $10 \, M_\odot$ to $\sim 300\, M_\odot$.

\section{Numerical Approach}

The evolution of low- and intermediate-mass primordial protostars has been
followed by Stahler, Palla, \& Salpeter (1986, hereafter SPS).  In that
study, an initially small core of 0.01 $M_{\sun}$ grows rapidly by accretion
at a rate of $\dot M=4.4 \times 10^{-3}$~M$_\odot$~yr$^{-1}$, corresponding
to the ambient gas temperature.  As shown by Omukai \& Nishi (1998), this is
consistent with the physical conditions in cloud fragments at the time of
core formation.  The calculation of SPS was terminated at a mass
$M_\ast=$10.5 M$_{\sun}$, when nuclear burning had not yet started and 
the core was well
below the mass scale of the parent cloud.  At that point, the protostar
was undergoing an internal readjustment due to strong heat transfer,
resulting in the appearance of a luminosity wave at the surface and radius
swelling.

The basic strategy and equations used here are the same as in SPS.  In this
scheme, the protostellar evolution is treated as a sequence of steady state
accretion flows onto a growing hydrostatic core (see Fig.1 of SPS for a
definition of the various regions). The core is assumed to be
in hydrostatic equilibrium and the ordinary stellar structure equations are
applied.  If the gas in front of the accretion shock is optically thin, no
envelope model is constructed and the boundary condition eq. (4a) of SPS is
applied at the core surface. In case the preshock gas is optically thick, we
integrate the equations inside the radiative precursor from the photosphere
to the core surface (see eqs. 7a-d of SPS).  Outside the photosphere, we
assume a free-falling flow and evaluate at each time the deceleration exerted
by radiation pressure.  
We adopted the same mass accretion rate of 
$\dot{M}=4.4 \times 10^{-3} \, M_{\sun}\, {\rm yr}^{-1}$ as in SPS.  
The main difference with that work is in the use of updated opacity tables: for
$T<7000$~K, we take the results of Lenzuni, Chernoff, \& Salpeter (1991,
composition of $X=72, Y=28$), and the OPAL opacity at higher temperatures
(Iglesias \& Rogers 1996, with slightly different composition $X=70, Y=30$). 

\section{Results}
The evolution of primordial protostars in the range $0.01M_{\sun}$ to
$10.5M_{\sun}$ has been computed by SPS.  According to their results, early
in the evolution, the accreted material settles adiabatically.  Because of
the high value of $\dot M$, the core is surrounded by an optically thick
radiative precursor.  As the protostar grows, the luminosity increases until
the Kelvin-Helmholtz (KH) time scale becomes shorter than the accretion
time.  The contraction causes the outward propagation of the accumulated heat
as a luminosity wave, which reaches the surface at a core mass of $M_\ast\sim
8 \,M_{\sun}$.  As a result, the radius starts increasing, and the radiative
precursor disappears.  Owing to the complexity of the internal readjustment
caused by the luminosity wave, we decided to follow again this phase and
chose the $M_\ast\simeq 8 \,M_{\sun}$ protostar of SPS as the initial model.

\subsection{Overall evolution}
The evolution of the core ($R_{\ast}$) and photospheric ($R_{\rm ph}$) radii
is presented in Figure 1.  Initially, the expansion of the core continues
from $M_\ast=$8 to 11.5 $M_{\sun}$, reaching a maximum value of $220
\,R_{\sun}$.  Then, due to the very short KH time, the protostar starts
contracting, and the radiative precursor appears again at a mass of
$M_\ast=12.4 ~M_{\sun}$.  Therefore, except for the short interval between
$M_\ast\sim 8$ and $12\, M_{\sun}$, the optically thick precursor persists
throughout the main accretion phase.  Unlike the core radius, the
photospheric radius steadily expands as the photospheric
luminosity ($L_{\rm ph}$) increases (see Fig. 2).  The spatial extent of the
precursor roughly corresponds to that of the ionized region because the
photospheric temperature, $T_{\rm eff} \simeq 6000$~K, is close to the
ionization temperature.  However, the present case is different from an
ordinary HII region in that here the gas is optically thick to electron
scattering and H$^{-}$ bound-free absorption, in addition to Lyman
continuum.

The evolution of the interior ($L_\ast$) and photospheric
luminosities divided by the core mass is shown in Figure 2.  The photospheric
luminosity is the sum of $L_\ast$ and $L_{\rm acc}$, the accretion luminosity.
As the luminosity-to-mass ratio $L/M_{\ast}$ grows, the effect of radiation
force increases relative to that of gravity.  Also, the accretion luminosity
$L_{\rm acc} \simeq GM_{\ast}\dot{M}/R_{\ast}$ increases as a result of core
contraction.  Because of this effect and the high accretion rate, $L_{\rm
ph}$ approaches the Eddington limit ($L_{\rm Edd}$) at $M_\ast\sim 60
\,M_{\sun}$.  At this point, the core radius has shrunk to $R_\ast \sim 12
\,R_{\sun}$, a value which is still higher than that of a zero-age main
sequence star ($2.8\,R_{\sun}$ for a $50\, M_{\sun}$ star, Marigo et al.
2001).

Despite the large increase of the interior luminosity, accretion is not
halted by the effect of radiation pressure.  What happens instead is a second
phase of expansion of the surface regions of the core starting at $M_\ast
\sim 60\, M_{\sun}$ (see Fig.~1) and a concurrent reduction of the accretion
luminosity.  The reason for the rapid increase of the radius is the
following.  While the inner part of the core continues contracting (see the
curve of $R_\ast$(90\%) in Fig. 1), the interior luminosity $L_{\ast}$ rises
roughly as $M_{\ast}^{2}$ and becomes close to the Eddington limit at
$M_\ast\sim 80~M_{\sun}$.  At this time, hydrogen burning begins and core
contraction stops. Owing to the high interior luminosity, strong radiation
forces act on the thin surface layers, where the opacity sharply peaks due to
the ionization of hydrogen and helium.  This causes an acceleration of the
expansion of the core from $\sim 80\, M_{\sun}$ to $\sim 100\, M_{\sun}$.
Then, the ratio $L_\ast/M_\ast$ levels off at a value close to, but still
smaller than, the Eddington limit (see Fig. 2) and the radius stops
increasing.

From then on, the surface layers of more massive protostars are subject to a
delicate balance between the radiation force and the external pressure of the
accreting envelope, whereas the interior relaxes toward the structure
appropriate to a main sequence star (see the evolution of $R_\ast (90\%)$ in
Fig. 1).  The interior luminosity is always close to the Eddington limit and
then increases approximately as $M_{\ast}$, although the ratio
$L_{\ast}/M_{\ast}$ keeps increasing slowly.  After further shrinking of the
core (between 100 and 260 $M_{\sun}$) , the photospheric luminosity
eventually reaches the Eddington limit.  The physical conditions in the
surface layers are thus similar to those that occurred at the time of the
first expansion, but now the course of events is much more dramatic.  The
interior luminosity $L_{\ast}$ continues to increase and eventually equals
the Eddington limit at $\sim 300\, M_{\sun}$.  This causes the huge expansion
of the outermost layers of the protostar shown in Fig. 1, while the interior
does not contract anymore.  The core runaway signals the end of the accretion
phase.

\subsection{Evolution of the Core Interior}
At the beginning of the calculations, the temperature is highest off-center
(around $q\equiv M/M_\ast \sim 0.1$) and the innermost part of the core
remains adiabatic.  In addition to this thermally inactive part, the
core consists of a radiative region and a superadiabatic surface layer.  The
newly accreted material is first incorporated into the superadiabatic layer,
where the material is heated by radiation.  In the radiative relaxation
layer, the material loses entropy and contracts.  

Because of the off-center temperature maximum, deuterium burning starts
off-center at about 12~$M_{\sun}$.  The deuterium luminosity reaches a
maximum between 12 and 16~$M_{\sun}$, but the contribution to the total
luminosity is always negligible ($<$10\%). Therefore, unlike Population I
and II protostars, in primordial objects deuterium burning does not play an
important role.  The central inert part persists until the stellar mass
reaches $\sim 40 M_{\sun}$, when radiative heating from outside becomes
strong enough and the off-center temperature maximum disappears.  We can see
the corresponding sudden rise (drop) of the central temperature (density) in
Figure 3b.  Thereafter, the central temperature continues to rise gradually
due to KH contraction.

When the core mass reaches $\sim 80\, M_{\sun}$, the central temperature exceeds
$10^{8}$K, needed to ignite the 3$\alpha$ reactions. Then, enough carbon is
built up to drive the CN cycle and significant hydrogen burning ensues (see
the sharp rise of $L_{\rm nuc}$ in Figure 2).  The onset and spread of
central convection is shown in Figure 3a.  Although carbon continues to be
synthesized at the center, the growth of the convective core and mixing
limits its abundance to values $\sim 10^{-9}$ (see Figure 3b).

Before the onset of the CN-cycle, the luminosity is generated mostly by
contraction, with a minor contribution of H-burning via the pp-chain.  This
situation differs from the standard ZAMS models studied by several groups who
find that the CN-cycle dominates the energy budget above $\sim 30 M_{\sun}$
(e.g. Ezer \& Cameron 1971).  Since the high entropy of the accreted matter
prevents contraction, the critical mass for the CN-cycle is much higher in
the protostellar case.

The radiation flow is absorbed by the high opacity in the superadiabatic
layer.  Because of the accumulated radiation energy in these regions,
radiation pressure exceeds gas pressure at $M_\ast \sim 60\, M_{\sun}$.  The
radiation dominated region quickly expands as the core mass increases.  On
the other hand, in the central convective core, the entropy increases in time
due to nuclear burning.  As a result, the central temperature remains almost
constant, whereas the density decreases gradually (see Figure 3b). The ratio
of radiation to total (gas$+$radiation) pressure keeps increasing, and
eventually the central part becomes dominated by radiation at
$M_\ast=145~M_{\sun}$.  This region expands rapidly and merges with the outer
radiation dominated region at $180~M_{\sun}$.  In more massive objects, the
internal structure is fully determined by radiation pressure.

\section{Discussion}

As a result of the high value of the mass accretion rate characteristic of
primordial gas clouds, the structural
properties of massive protostars have several unique and unexpected features.  
The evolutionary tracks of the photosphere and core surface in the HR diagram
are displayed in Figure 4.  The locus of the population III ZAMS stars
computed from static models is also shown (Marigo et al. 2001; Bromm et al.
2001).  The effective temperature of the protostellar core is computed from
the core radius and preshock luminosity.  The striking feature of the
photospheric track is the almost constant value of the effective
temperature.  The photosphere forms in the accreting envelope, and persists
almost throughout the main accretion phase.  Because of the 
sensitivity of the H$^{-}$ bound-free opacity, the temperature is
locked at the value of $T_{\rm eff}\simeq 6000$~K (see eq. 23b of SPS). 
Observationally, primordial
protostars of vastly different luminosities should have the same optical
colors.  On the other hand, the core surface will remain unobservable from
outside because of the presence of the optically thick precursor.

From Fig.~4 we see that the track of the core surface cuts through the ZAMS
track of the core surface at a mass $\sim 40~M_\sun$.  However, the curve does
not join the ZAMS smoothly, but follows a more elaborate path due to the
behavior of the mass-radius relation. Therefore, realistic ZAMS models of
Pop~III stars should start with initial conditions that differ from those
usually adopted on the basis of thermal equilibrium.

Also, the nuclear properties derived from the protostellar phase are quite
distinctive.  Deuterium burning and the p-p chain reactions have little
effect on the evolution.  Hydrogen burning due to the CN-cycle begins only at
$\sim 80~M_{\sun}$.  At this mass, the interior luminosity becomes close to
the Eddington limit, so that the surface regions are subject to strong
radiation forces.  These conditions are favorable to the onset of activity
in the form of mass loss. It is clear that continuum 
radiation driven winds can occur in Pop III protostars. According
to the mass-radius relation of Fig.~1, there are two critical stages 
when the conditions become favorable for the onset of a wind: the
rapid expansion of the surface layers at $M_\ast \sim 80$ and $300\, M_\sun$.
However, it is only at the highest mass that we expect a major
episode of mass loss, since earlier on the interior luminosity always
remains slightly below the critical value. Even in this case, however,
we anticipate a short lived phase of wind activity related to the 
small extent of the high opacity surface layer.

Our result of a upper mass limit of $\sim 300\, M_\sun$ is sensitive to the
adopted value of the mass accretion rate and to its time evolution. 
In fact, $\dot M$ may not be constant and decrease in time from an even
larger initial value (e.g., Omukai \& Nishi 1998).
Quantitatively, it is difficult to
predict the exact values of $M_\ast^{\rm max}$ as a function of $\dot M$
without detailed models. 
However, we can expect that the evolution obtained
here will not change qualitatively. In case of larger values of $\dot M$, the
higher luminosity will cause an earlier stripping of the surface layers by
radiation pressure.  Also, nuclear burning will be postponed at higher masses
because of the shorter contraction time scale.  On the contrary, for lower
accretion rates, the evolution would resemble that of Pop I protostars with
the central core smoothly becoming a ZAMS star during the main accretion
phase, but with a fundamental difference.
Because of the lack of grains, the radiative force is not capable to reverse 
the infall until a very large value of the protostellar mass. 
Thus, it is evident that there must be a {\it critical} value of
$\dot M$ for the onset of the runaway increase of the protostellar radius
that determines the maximum accumulated mass. 
The existence of this critical
behavior will be explored in a forthcoming paper.

\acknowledgements This work is supported in part by Research Fellowships of
the Japan Society for the Promotion of Science for Young Scientists, grant
6819. K.O. wishes to thank the director of the Osservatorio di Arcetri for
the hospitality during an extended visit in 2001.


\plotone{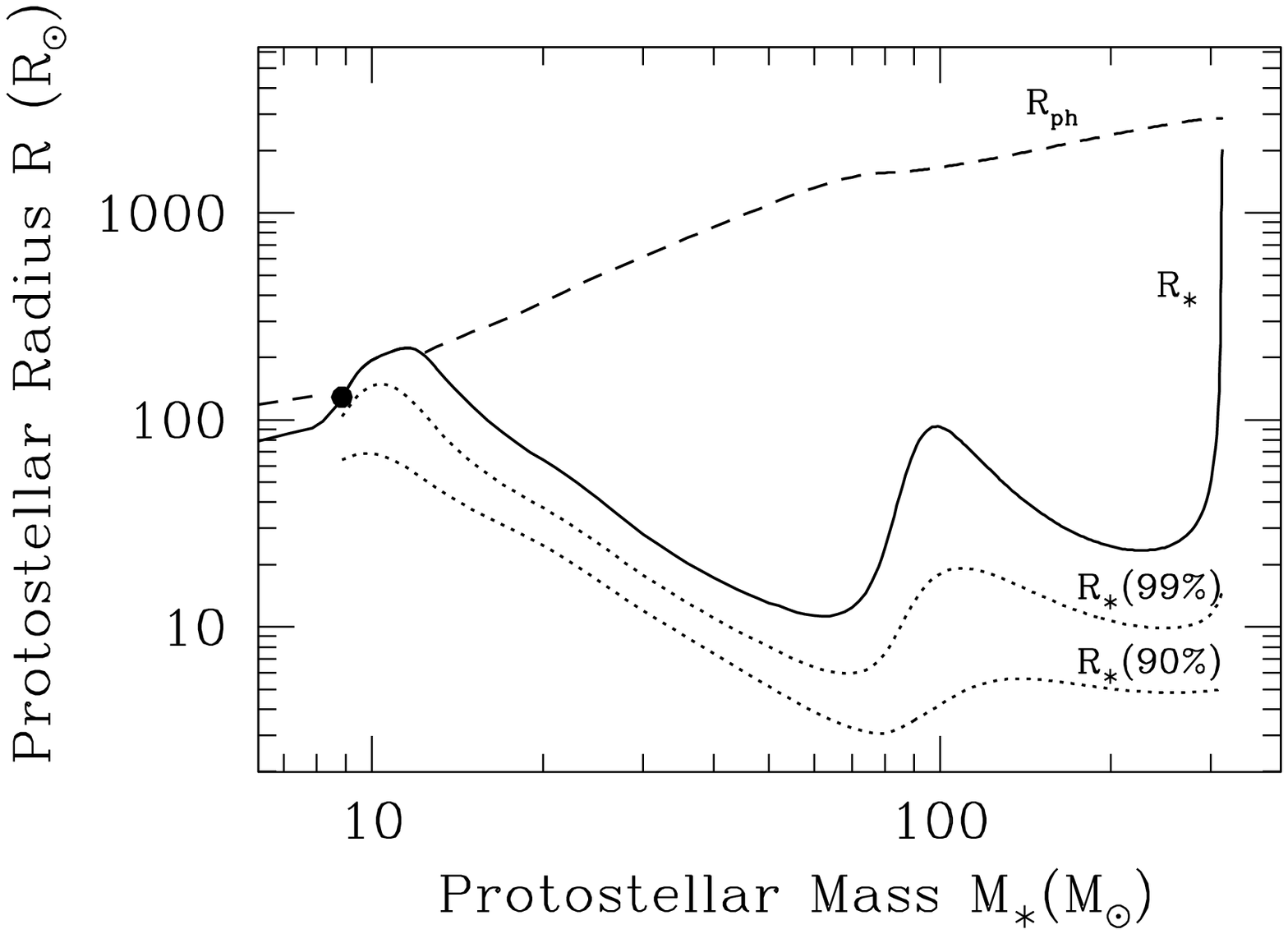}
\figcaption[f1.eps]{Mass-radius relation for massive primordial
protostars. The evolution of the core (solid line) and photospheric 
(dashed line) radii is shown as a function of the protostellar mass.
The circle on the curve of $R_\ast$ marks the initial model
of our calculations. The $R_\ast$ and $R_{\rm ph}$ curves 
at smaller masses come from SPS.
The dotted lines represent the radii of mass shells containing 
90\% and 99\%, respectively, of the total stellar mass. 
\label{fig:rad}}

\plotone{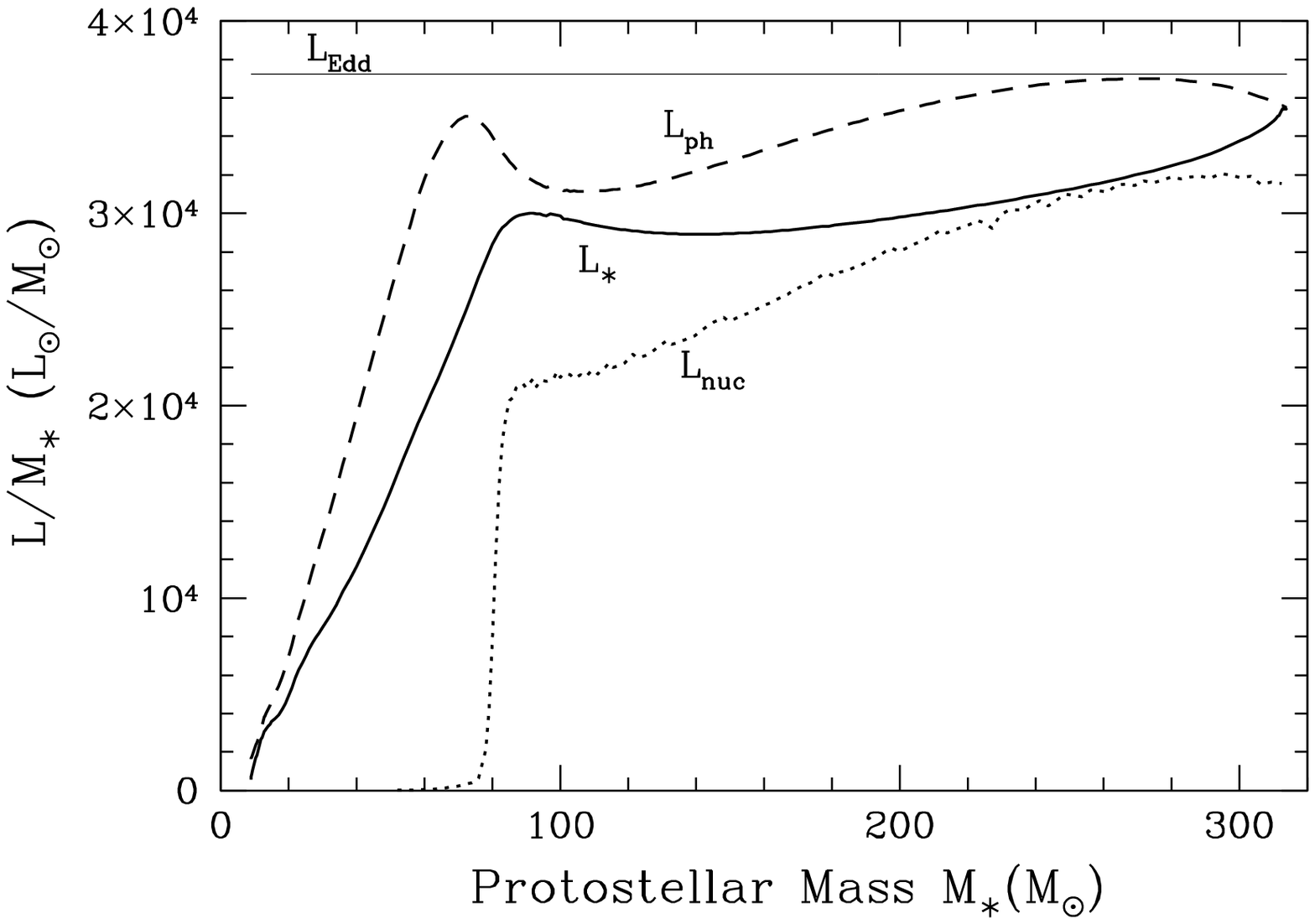}
\figcaption[f2.eps]{Evolution of the luminosity to mass ratio as a function 
of mass.  The interior and photospheric luminosities are displayed by the
solid and dotted lines, respectively.
The solid horizontal line represents the Eddington luminosity
for electron scattering. The nuclear luminosity due to H-burning is
shown by the dashed line.
\label{fig:Lum}}

\plotone{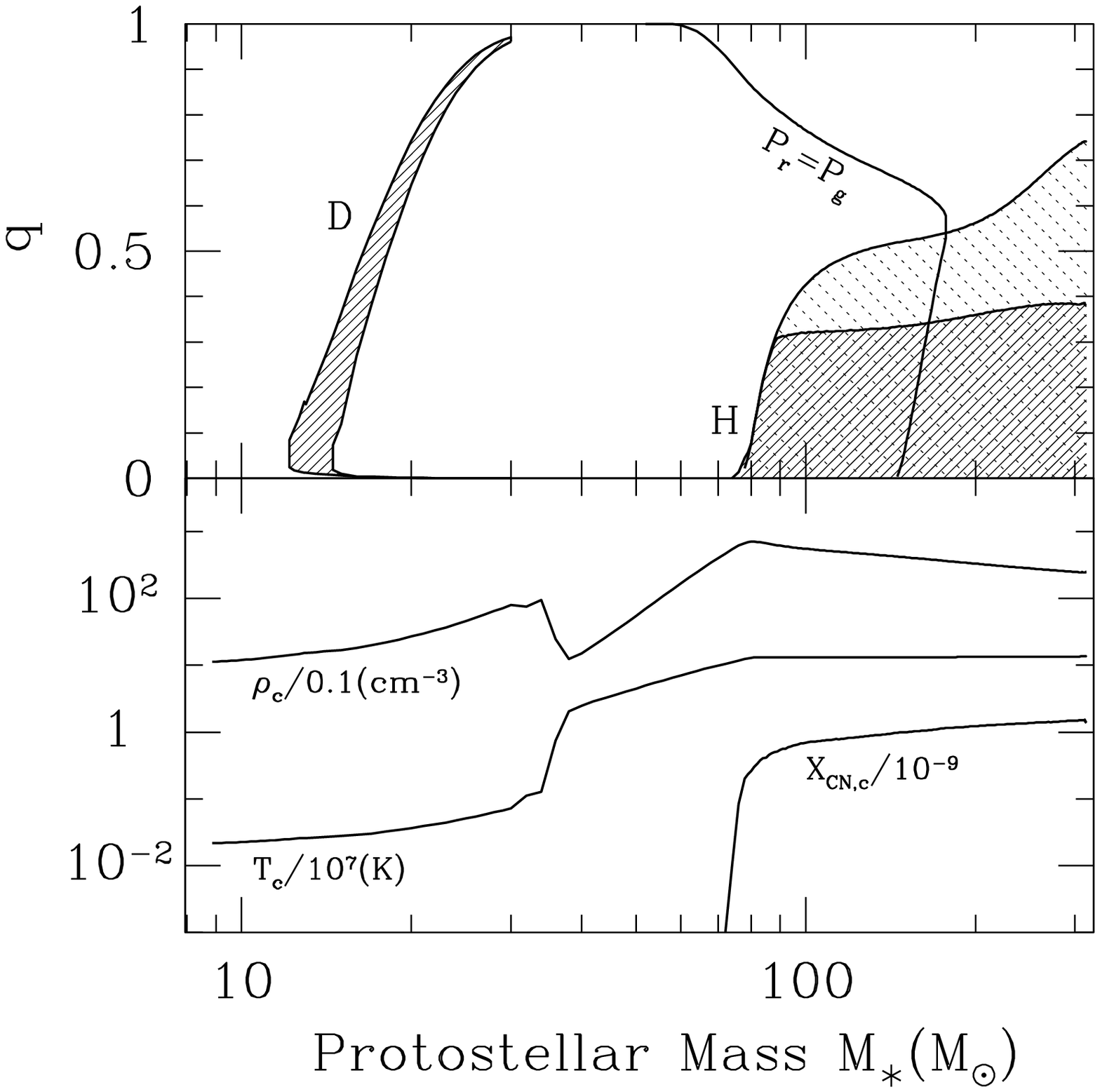}
\figcaption[f3.eps]{The internal structure of the core as a function of mass.
(a) Evolution of the nuclear energy and convection as a function
of the relative mass $q \equiv M/M_{\ast}$.
Regions where the energy generation by D- and H-burning 
exceeds 10\% of the average energy generation rate $L_{\ast}/M_{\ast}$ 
are shown as shaded regions
The extent of the convective core is illustrated by the short-dashed area. 
The curve labelled $P_{\rm r}=P_{\rm g}$ is the locus where the radiaton
pressure equals gas pressure. Radiation pressure dominates
to the right of this curve.
(b) Evolution of central density, temperature, and mass fraction of 
CN-elements.
Each variable is normalized to $0.1~{\rm g~ cm^{-3}}$, $10^{7}$~K, 
and $10^{-9}$, respectively.
\label{fig:core}}

\plotone{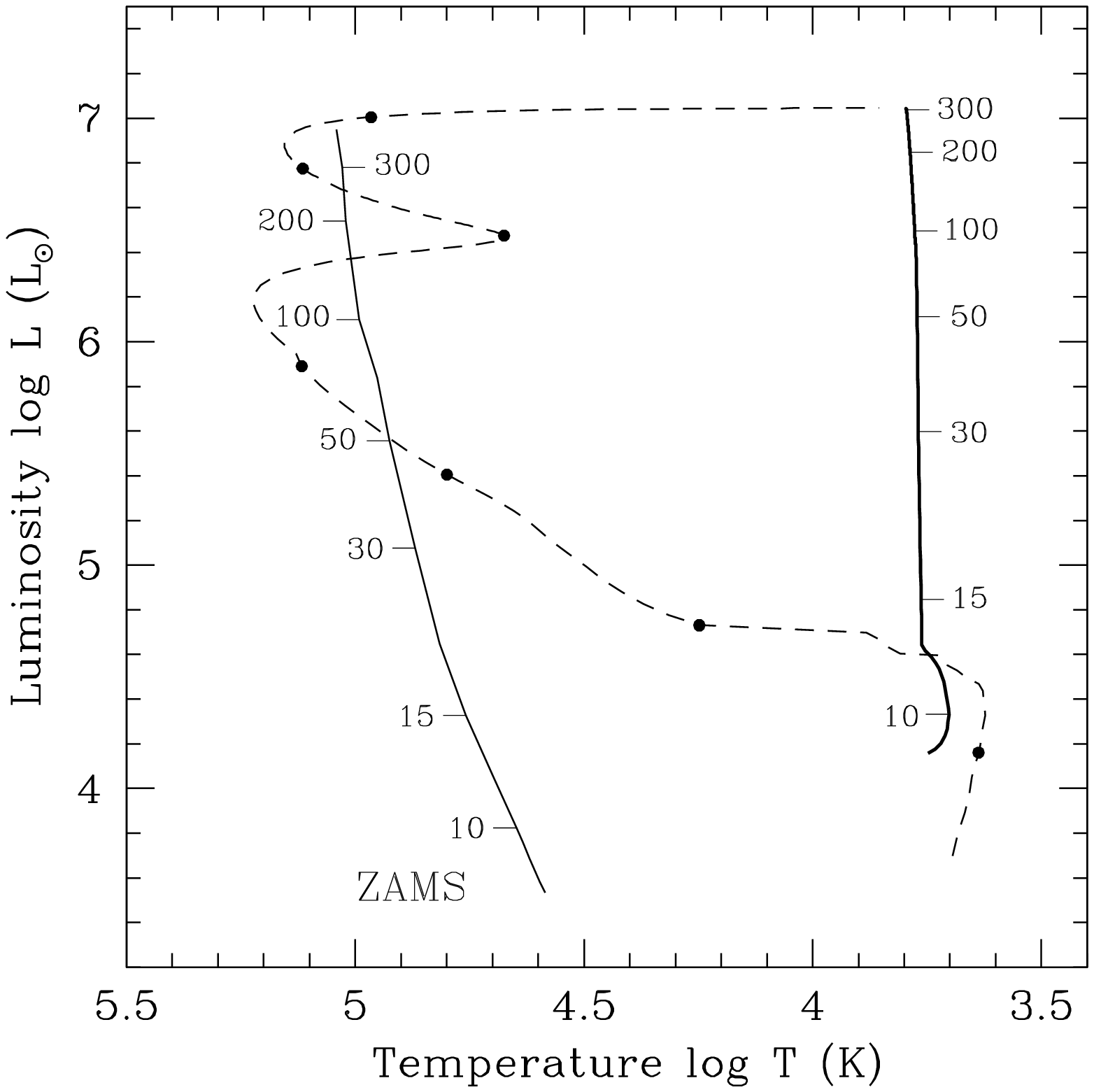}
\figcaption[f4.eps]{HR diagram for primordial protostars.
The evolution of the photosphere and core surface is shown by the 
thick solid and dashed lines, respectively. 
For comparison, we also show the
locus of the metal-free ZAMS stars 
(Marigo et al. 2001 for $M_\ast<100\, M_{\sun}$; 
Bromm, Kudritzki, \& Loeb 2001 for higher mass).
The numbers on both tracks label the value of the core mass (in solar units).
The filled circles on the dashed line have the same meaning.
\label{fig:HR}}

\end{document}